\documentclass[11pt,twoside]{article}
\usepackage{asp2010}
\usepackage{graphicx}
\usepackage{natbib}

\resetcounters

\begin{document}

\title{Dust grain growth in the interstellar medium of galaxies at redshifts $4<z<6.5$}
\author{Micha{\l}~J.~Micha{\l}owski$^1$, 
Eric~J.~Murphy$^2$,
Jens~Hjorth$^3$,
Darach~Watson$^3$,
Christa~Gall$^3$,
and
James~S.~Dunlop$^1$
\affil{$^1$Scottish Universities Physics Alliance, Institute for Astronomy, University of Edinburgh, Royal Observatory, Edinburgh, EH9 3HJ, UK}
\affil{$^2$Spitzer Science Center, MC 314-6, California Institute of Technology, Pasadena, CA 91125, USA}
\affil{$^3$Dark Cosmology Centre, Niels Bohr Institute, University of Copenhagen, Juliane Maries Vej 30, 2100 Copenhagen \O, Denmark}
}

\begin{abstract}
To discriminate between  different dust formation processes is a key issue in order to understand its properties. We analysed six submillimeter galaxies at redshifts $4<z<5$ and nine quasars at $5<z<6.4$. We estimated their dust masses from their (sub)millimeter emission and their stellar masses from the spectral energy distribution modelling or from the dynamical and gas masses obtained from the CO line detections. We calculated the dust yields per AGB star and per SN required to explain these dust masses and concluded that AGB stars are not efficient enough to form dust in the majority of these galaxies. SN could be responsible for dust production, but only if dust destruction in the SN shocks is not taken into account. Otherwise even SNe are not efficient enough, which advocates for some other dust production mechanism. We present the hypothesis that grain growth in the interstellar medium is responsible for bulk of the dust mass accumulation in these galaxies.

\end{abstract}

\section{Introduction}

Dust is of prime importance in cosmology, not only because it obscures our view on stellar populations, but also because its emission contains crucial information about the half of the energy emitted in the Universe. Dust can be formed by asymptotic giant branch (AGB) stars and by supernovae (SNe).

It is known that an AGB star can produce up to $\sim4\mbox{}\times10^{-2}\,M_\odot$ of dust \citep{morgan03,ferrarotti06}, whereas a SN can produce up to $\sim1.32\,M_\odot$ of dust \citep{todini01,nozawa03}. However, the SN shocks destroy the majority of the dust formed by a SN leaving only $\lesssim0.1\,M_\odot$  to be ejected into the interstellar medium  \citep[ISM;][]{bianchi07,cherchneff10}. 

The dust yields derived observationally for SN remnants are usually $<0.01\,M_\odot$ except of Cassiopeia~A \citep{dunne03,dunne09casA} and Kepler \citep{morgan03b,gomez09}, for which the derived dust yields are of the order of $\sim1\,M_\odot$.

It is difficult to distinguish the dust formed by AGB stars and by SNe. Based on the flat extinction curves, the SN-origin dust has been claimed to be present in distant quasars (QSOs) and gamma-ray burst host galaxies (\citealt{maiolino04}, \citealt{gallerani10}, \citealt{perley10} and \citealt{stratta07b}, but see \citealt{zafar10}).

Another option is that the stellar sources provide only the dust seeds and the bulk of the dust mass accumulation happens in the ISM by the grain growth via capturing of heavy elements  \citep{draine79,dwek80,draine90,draine09}.

The significant amounts of dust have been detected in the early Universe up to $z\sim6.4$ \citep{dunlop94,benford99,archibald01,omont01,priddey01,priddey03,priddey08,isaak02,bertoldi03,robson04,beelen06,wang08b,martinezsansigre09,michalowski10smg,santini10}. Hence, its origin has to be explained by a process that is both efficient and rapid.

Several works have attempted modelling of the dust evolution at these high redshifts  \citep{dwek07,valiante09,dwek11,gall11,gall11b,pipino11}. In this paper we present more simplistic approach of estimating whether stellar sources are efficient enough to explain the dust detected at high redshifts.

\section{Method}

We selected six submillimeter galaxies (SMGs) at $4<z<5$ \citep[][note that the currently highest-redshift SMG from \citealt{riechers10} and \citealt{capak11} is not included in this study]{coppin09,capak08,schinnerer08,daddi09,daddi09b,knudsen09}  and modelled their spectral energy distributions (SEDs) using the radiative transfer code GRASIL \citep{silva98} utilising the  35\,000 SED templates of \citet{iglesias07}
The details of the SED modelling can be found in \citet{michalowski08,michalowski09,michalowski10smg,michalowski10smg4, michalowski11}.

To extend the explored redshift range we selected nine quasars (QSOs) at $5<z<6.4$. Their near-IR emission is dominated by active galactic nuclei instead of stellar population, so we approximated their stellar masses as a difference between the dynamical and the gas masses derived from the CO line detections \citep{walter04,maiolino07, wang10}.

The number of stars with masses between $M_0$ and $M_1$ in the stellar population with a total mass of $M_*$ was calculated as $N(M_0$--$M_1)=M_* \int_{M_0}^{M_1} M^{-\alpha} dM / \int_{M_{\rm min}}^{M_{\rm max}} M^{-\alpha}  M dM$. 
It is unclear what kind of the initial mass function (IMF) is appropriate at high redshifts  \citep{baugh05,fontanot07,hayward11, hayward11b}.
Hence, we adopted an IMF with $M_{\rm min}=0.15$, $M_{\rm max}=120\,M_\odot$, and a slope $\alpha=2.35$ \citep[][or $\alpha=1.5$ for a top-heavy IMF]{salpeter}. 
The average dust yield per  star is $M_{\rm dust} /  N(M_0$--$M_1)$.

Stars with masses $<8\,M_\odot$ need at least $50$ Myr to enter the AGB phase and start producing dust, so for AGB stars the $M_*$ was replaced with the $M_*-M_{\rm burst}$, where $M_{\rm burst}$ is the mass of stars created in the last $50$ Myr.

\section{Results and discussion}

\begin{figure}[!ht]
\begin{center}
\includegraphics[width=\textwidth]{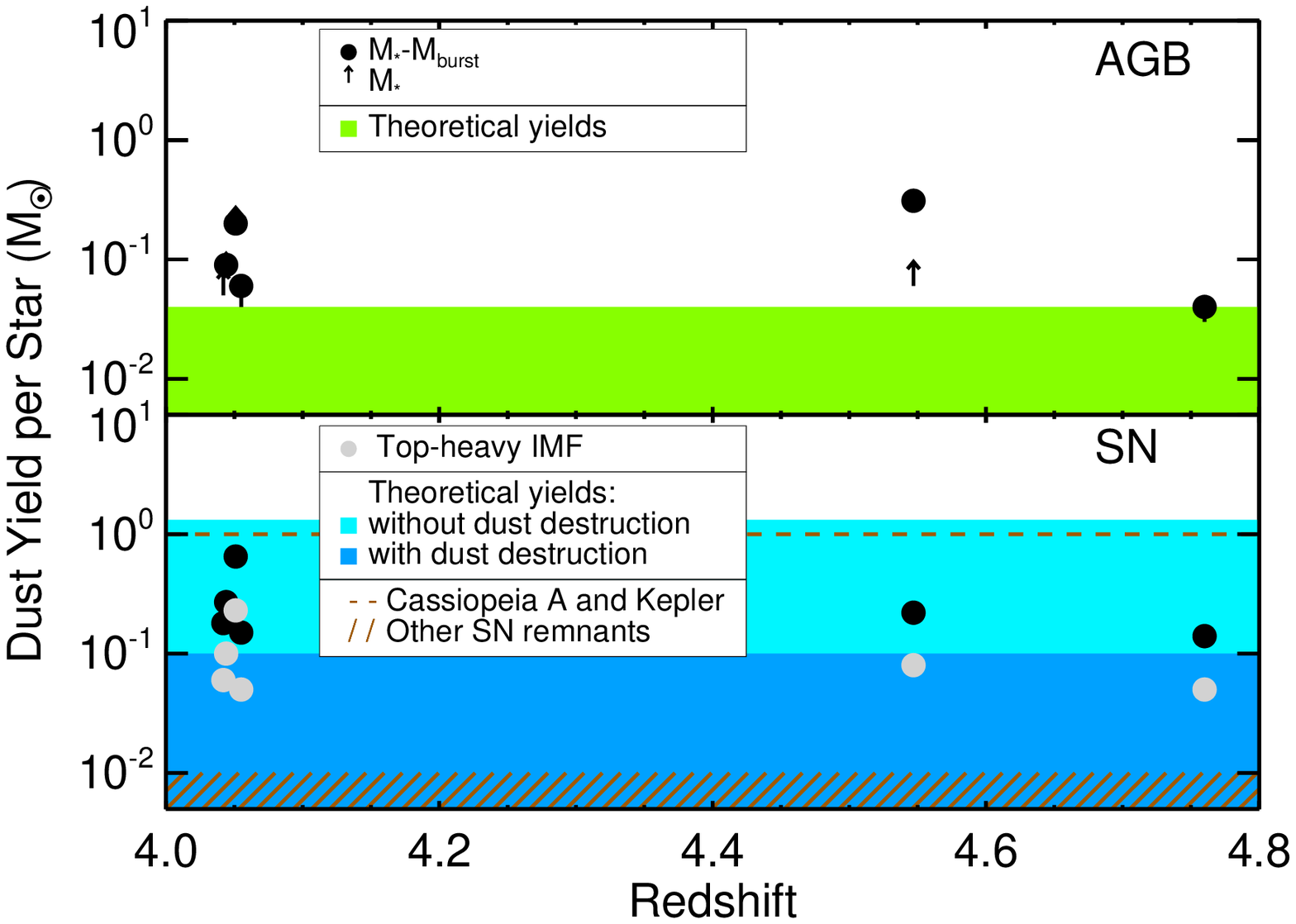}
\end{center}
\caption{Dust yields per AGB star ({\it top}) or per SN ({\it bottom}) required to explain dust in the $4<z<5$ SMGs \citep{michalowski10smg4}.  The extension to $z > 5$ is presented in \citet{michalowski10qso}.
AGB stars are not efficient enough  and SNe would need to be unfeasibly efficient to form dust in these sources suggesting rapid grain grown in the ISM is likely to be responsible for the large dust masses.
{\it Circles}: the best estimates of the required dust yields.
{\it Arrows}: lower limits calculated for stellar masses without subtracting the young stars formed in the last $50$ Myr.
{\it Gray symbols} indicate that a top-heavy IMF was adopted.
{\it Dashed line} and {\it diagonal lines}: the dust yields derived for  Cassiopeia A, 
Kepler
($\sim1\,M_\odot$)
and other SN remnants 
($\sim10^{-3}$--$10^{-2}\,M_\odot$),
respectively. 
{\it Green area}: theoretical dust yields for AGB stars 
($\lesssim4\cdot10^{-2}\,M_\odot$).
{\it Light blue} and {\it blue areas}: theoretical SN dust yields without 
($\lesssim1.32\,M_\odot$)
and with dust destruction implemented 
($\lesssim0.1\,M_\odot$), respectively.}
\label{fig:dust}
\end{figure}

The required dust yields for SMGs \citep{michalowski10smg4} are shown on Figure~\ref{fig:dust} as a function of redshift. For similar representation of required dust yields for QSOs see Figure~1 of \citet{michalowski10qso}.

For both SMGs and QSOs the AGB stars are not efficient enough to form dust in these objects as their required dust yields are higher than the range allowed by the theoretical modelling.

If SNe do not destroy the dust they form (i.e.~eject up to $\sim1\,M_\odot$ of dust into the ISM) then they are efficient enough to explain dust in all $z>4$ SMGs and QSOs discussed here. However, the models predict that only $\sim10$\% of this dust survives the shocks \citep{bianchi07,cherchneff10} and then even SNe cannot explain the dust in the most of these galaxies.
 
 We checked that this conclusion does not change when we allow both AGB stars and SNe to form only a fraction of dust \citep[Fig.~2 of][]{michalowski10qso}. It is because the required yields for AGB stars exceed the allowed value by a factor of $>4$ in most of the cases and therefore AGB stars can account for only $<25$\% of dust detected in $z>5$ QSOs. Therefore the required dust yields for SNe can be scaled down only slightly which does not suffice to bring them into the consistency with models with dust destruction implemented.
 

Hence, we found that  neither AGB star nor SNe are able to fully account for dust detected in some $z>4$ galaxies. This indicates that a significant fraction of dust in these galaxies was formed by some other mechanism. One of the possibility is that the grain growth in the ISM is responsible for dust accumulation. 

This process is relatively fast with the timescale of a few$\mbox{}\times10$ Myr \citep{draine90,draine09,hirashita00,zhukovska08}. Therefore even at redshift $6.4$ this timescale is much shorter than the likely age of a galaxy. Moreover, SNe deliver enough heavy elements to fuel this growth \citep{todini01,nozawa03,bianchi07,cherchneff09}.

\section{Conclusions}

We investigated the possible dust producers in the SMGs at $4<z<5$ and QSOs at $5<z<6.4$ by estimating the required dust yields per AGB star and per SN and comparing them to the allowed values  derived from the theoretical models. We found that AGB stars are not efficient to form dust in the majority of these galaxies leaving SNe as the only likely stellar dust producers. However, the required dust yields for SNe exceed the allowed values if dust destruction in the SN shocks is taken into account. This advocates for some other dust production mechanism e.g.~the grain growth in the ISM. This process is fast enough and SNe deliver enough heavy elements to build up the dust masses derived for these galaxies.

\acknowledgements We  thank Joanna Baradziej for help with improving this paper. MM and JSD acknowledges the support of the Science and Technology Facilities Council. JSD acknowledges the support of the Royal Society via a Wolfson Research Merit award, and the support of the European Research Council via the award of an Advanced Grant.
The Dark Cosmology Centre is funded by the Danish National Research Foundation.

\end{document}